\documentclass[preprintnumbers,showkeys,byrevtex]{revtex4}
\usepackage{amsmath,amsfonts,amssymb,amscd,amsxtra,amsthm}
\usepackage{graphicx}
\usepackage{bm}
\usepackage{epstopdf}
\usepackage{multirow}
\begin{document}
\preprint{KIAS-P12049}
\title{Extended nonlocal chiral-quark model for the heavy-light quark systems}
\author{Seung-il Nam}
\email[E-mail: ]{sinam@kias.re.kr}
\affiliation{School of Physics, Korea Institute for Advanced Study (KIAS), Seoul 130-722, Republic of Korea}
\date{\today}
\begin{abstract}
In this talk, we report the recent progress on constructing a phenomenological effective model for the heavy-light quark systems, which consist of $(u,d,s,c,b)$ quarks, i.e. extended nonlocal chiral-quark model (ExNLChQM). We  compute the heavy-meson weak-decay constants to verify the validity of the model. From the numerical results, it turns out that $( f_D, f_B, f_{D_s}, f_{B_s})=(207.54,208.13,262.56,262.39)$ MeV. These values are in relatively good agreement with experimental data and various theoretical estimations.  
\end{abstract}
\keywords{Heavy-quark effective field theory, charm and bottom quarks, heavy mesons, instanton, nonlocal chiral-quark model }
\maketitle
\section{Introduction}
The heavy-quark effective field theory (HQEFT) is a very powerful theoretical tool to investigate various heavy hadrons~\cite{Georgi:1990um}. HQEFT manifests the heavy-quark symmetries: Spin and heavy-flavor symmetries~\cite{Georgi:1990um}. In general, they are not obvious in full QCD but become relevant in the heavy-quark limit $m_Q\to\infty$. Using HQEFT, the light-quark component of the heavy meson was treated nonperturbatively, manifesting the spontaneous  breakdown of chiral symmetry (SBCS), in many different methods~\cite{Ebert:2006hj,Badalian:2007km,Hwang:2010hw,Geng:2010df}. In the present talk, we report the recent progress on constructing a {\it new} effective model for the heavy hadrons. The model is based on two ingredients in principle: HQEFT and nonlocal chiral-quark model (NLChQM) for the heavy and light quarks, respectively~\cite{Praszalowicz:2001wy,Nam:2006au,Dorokhov:2010zzb}, resulting in the extended NLChQM (ExNLChQM). Intuitively, the heavy-meson mass can be regarded as the total sum of the various light and heavy quark masses, i.e. the effective and current ones:
\begin{equation}
\label{eq:MASS}
\underbrace{1869\,\mathrm{MeV}}_{D\,\mathrm{meson\,mass}}\hspace{0.7cm}>\underbrace{350\,\mathrm{MeV}}_\mathrm{light\,quark\,effective\,mass}+\underbrace{1290\,\mathrm{MeV}}_\mathrm{heavy\,quark\,current\, mass}.
\end{equation}
From Eq.~(\ref{eq:MASS}), one needs additional $200$ MeV mass. Hence, we assume that this additionally ({\it effective}) mass can be generated from the similar mechanism with the light quark, i.e. from the nontrivial contributions from the QCD vacuum. For more details, one can refer~\cite{Nam:2011ak,Nam:2012se}. This is a main idea to construct ExNLChQM. We organize the report as follows: In Section II, we discuss the formulation of ExNLChQM. Section III is devoted to the numerical results with relevant discussions. Summary and conclusion will be given in Section IV.

\section{Theoretical framework}
Using a generic functional integral technique, we can arrive at an effective action for the heavy-light quark systems from the effective Lagrangian density of ExNLChQM~\cite{Nam:2011ak,Nam:2012se}:
\begin{eqnarray}
\label{eq:EFA3}
\mathcal{S}^\mathrm{LL+HL+LH}_\mathrm{eff}=
-i\mathrm{Sp}\ln
\Bigg[-\left(\frac{1}{ f_H}\sqrt{M_Q}H\sqrt{M_q}
\right) (iv\cdot\partial-M_Q)^{-1}
\left(\frac{1}{ f_H}\sqrt{M_q}\bar{H}\sqrt{M_Q}\right)\Bigg],
\end{eqnarray}
where $\bar{M}_q=m_q+M_q$. The momentum-dependent effective masses for the heavy and light quarks are given by
\begin{equation}
\label{eq:MMM}
M_{(q,Q)}
=M_{(q,Q),0}\left[\frac{2\Lambda^2_{(q,Q)}}{2\Lambda^2_{(q,Q)}-|i\rlap{/}{\partial}|^2} \right]^{2},
\end{equation}
where $(\Lambda_q,\Lambda_Q)=(600,1000)$ MeV~\cite{Nam:2011ak,Nam:2012se}. $ f_H$ denotes the weak-decay constant for the heavy meson. We determine $M_{Q,0}$, which is the effective heavy-quark mass at zero virtuality, from a simple phenomenological analysis:
\begin{equation}
\label{eq:HMMM}
M_H\approx
\left[m_q+M_{q,0} \right]_\mathrm{L}+\left[m_Q+M_{Q,0} \right]_\mathrm{H}.
\end{equation}
From the experimental data for $D$ and $B$ mesons, we can estimate the values of $M_{Q,0}$ for numerical calculations with $M_{q,0}\approx350$ MeV~\cite{Nam:2006au}: 
\begin{eqnarray}
\label{eq:ES}
M_D&=&1869.57\,\mathrm{MeV}\approx1625.0\,\mathrm{MeV}+M_{Q,0}\to M_{Q,0}\approx 244.57\,\mathrm{MeV}
\cr
M_{B}&=&5279.17\,\mathrm{MeV}\approx5025.0\,\mathrm{MeV}+M_{Q,0}\to M_{Q,0}\approx 254.17\,\mathrm{MeV}.
\end{eqnarray}
The heavy-meson weak-decay constant, $ f_H$ is defined by the following:
\begin{equation}
\label{eq:FPI}
\langle0|\bar{q}(x)\gamma_\mu(1-\gamma_5)\tau^aQ_v(x)|H^b(p)\rangle
=2ip_\mu  f_H\delta^{ab}.
\end{equation}
The matrix element in the left-hand-side of Eq.~(\ref{eq:FPI}) can be easily derived from the functional derivative of the effective chiral action in Eq.~(\ref{eq:EFA3}). 

\section{Numerical results}
\begin{figure}[t]
\begin{tabular}{cc}
\includegraphics[width=6cm]{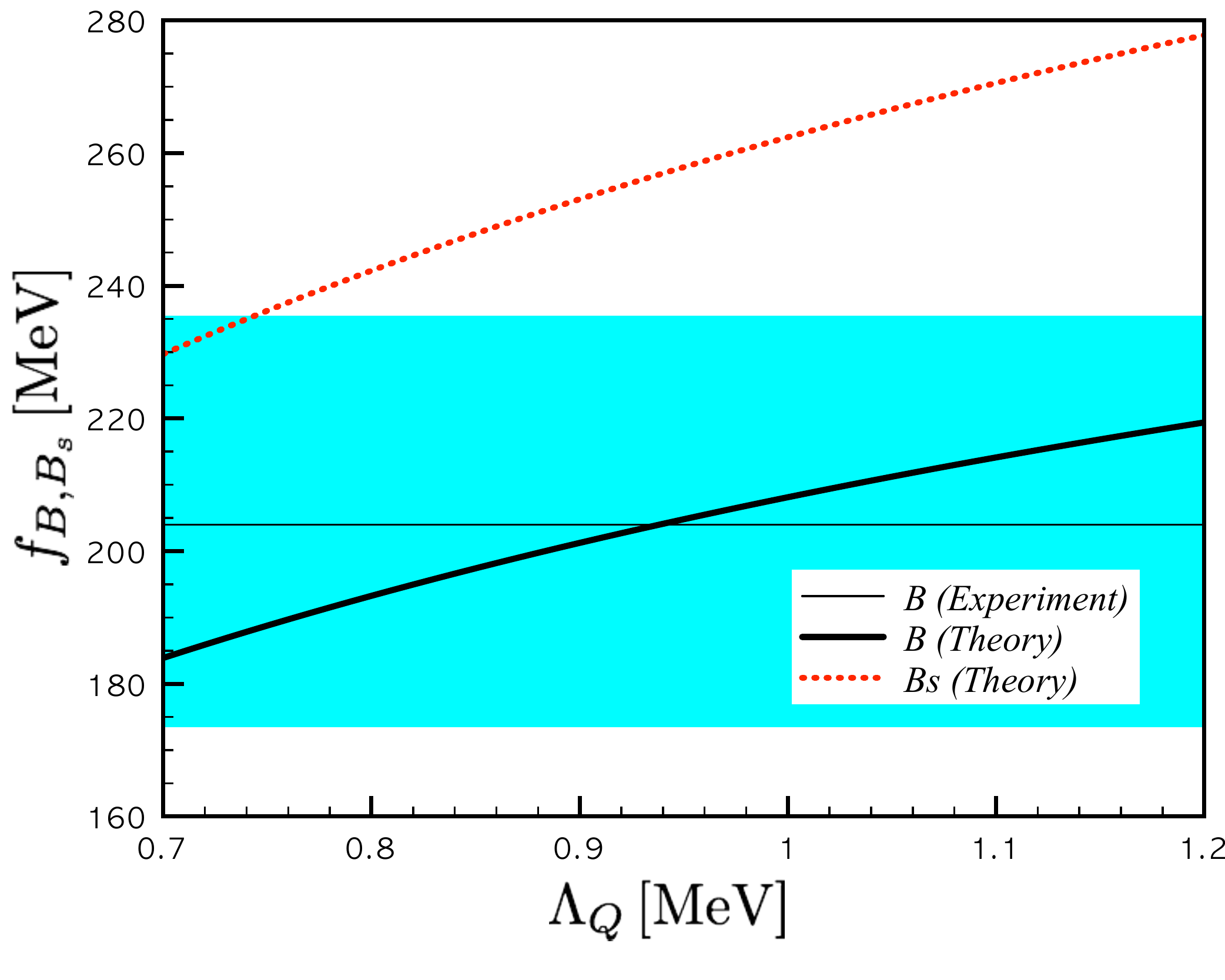}
\includegraphics[width=6cm]{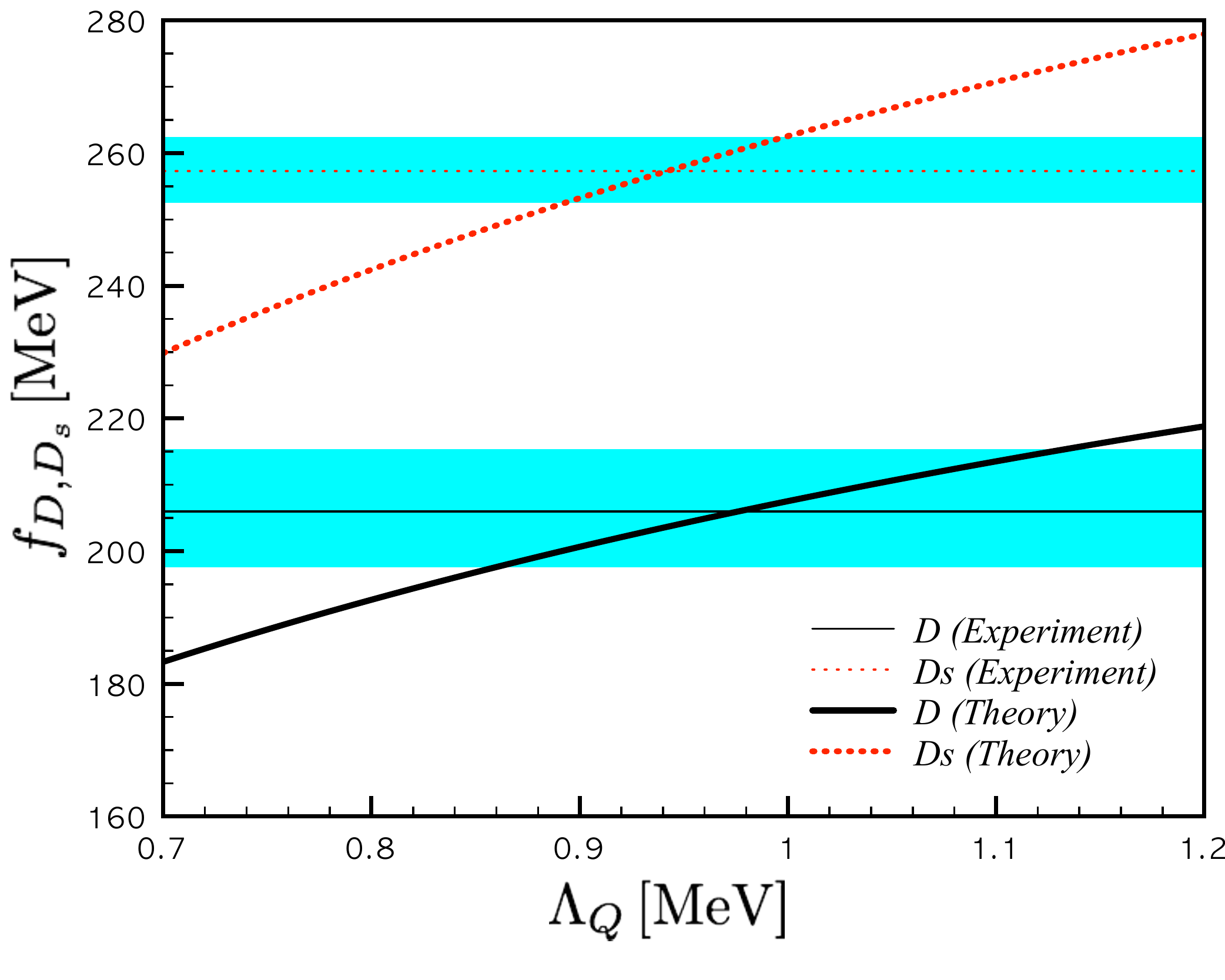}
\end{tabular}
\caption{(Color online) $ f_{D,D_s}$ (left) and $ f_{B,B_s}$ (right) as functions of $\Lambda_Q$ at $\Lambda_q=600$ MeV. The numerical results for the non-strange and strange mesons are shown with the solid and dot lines, respectively. The experimental data are taken from~\cite{Rosner:2008yu,Rosner:2010ak}. Note that the experimental errors are given with the shaded areas.}       
\label{FIG1}
\end{figure}
\begin{figure}[t]
\begin{tabular}{cc}
\includegraphics[width=6cm]{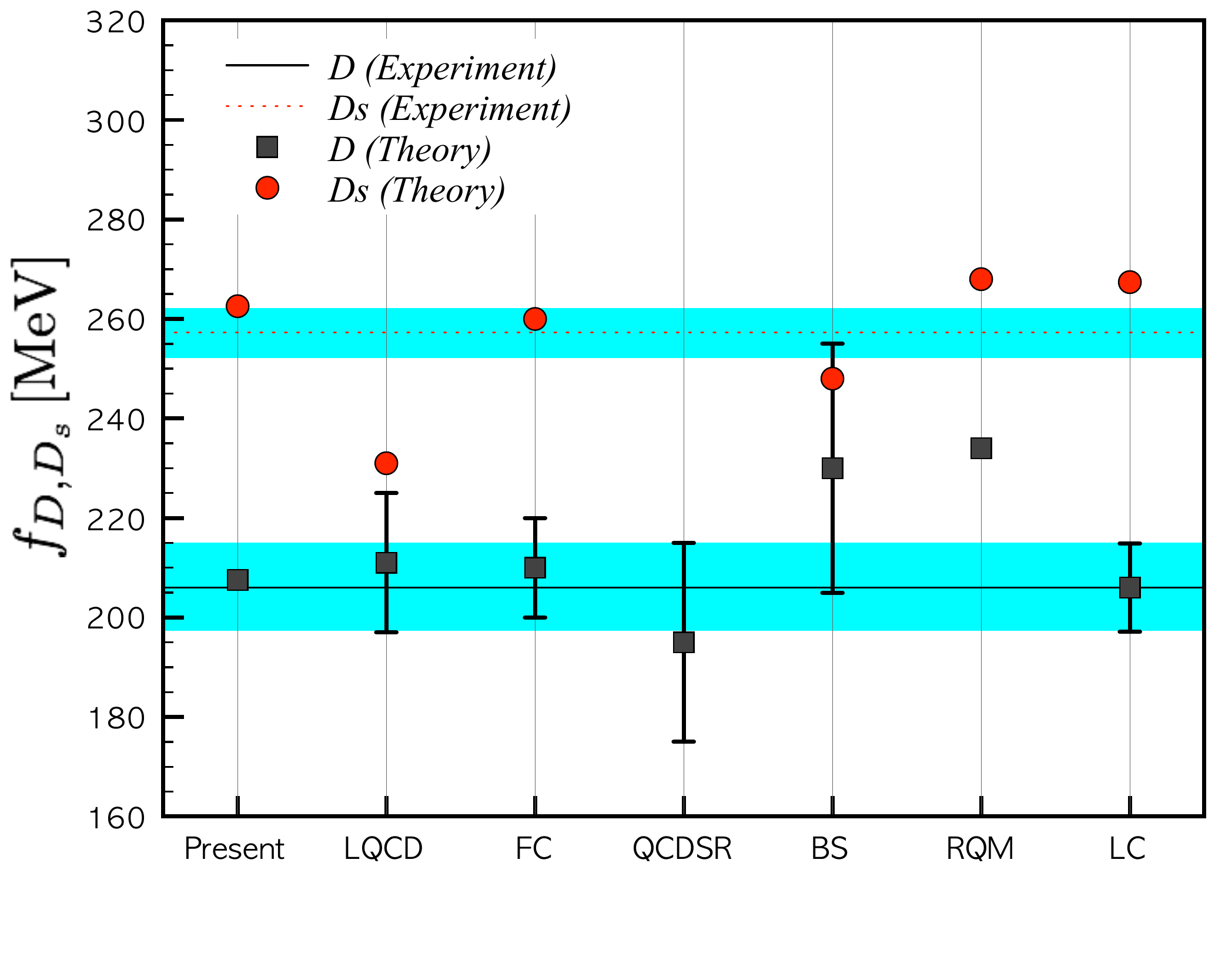}
\includegraphics[width=6cm]{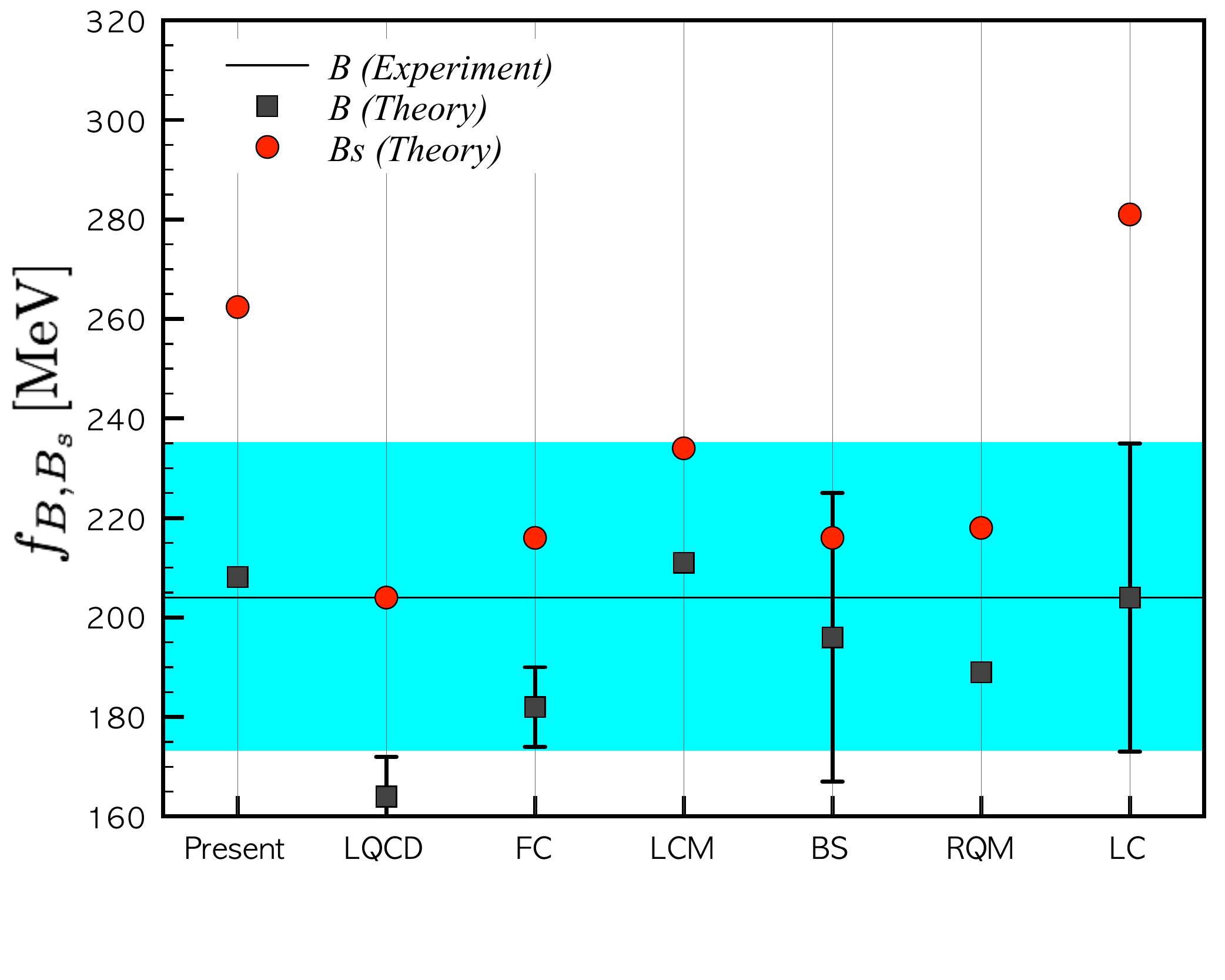}
\end{tabular}
\caption{(Color online) Theoretical results for $ f_{D,D_s,B,B_s}$ from the present calculations, clover-improved quenched LQCD~\cite{Becirevic:1998ua},  field-correlator method (FC)~\cite{Badalian:2007km},  QCD sum rule~\cite{Penin:2001ux}, light-front quark model (LQM)~\cite{Choi:2007se}, Bethe-Salpeter method (BS)~\cite{Cvetic:2004qg}, relativistic quark model (RQM)~\cite{Ebert:2006hj}, and light-cone wave function (LC)~\cite{Hwang:2010hw}.}       
\label{FIG2}
\end{figure}

In Fig.~\ref{FIG1}, we depict the numerical results for $ f_H$ as functions of $\Lambda_Q$ by fixing $\Lambda_q=600$ MeV. The horizontal lines indicate the global fit data with errors (shaded areas). To reproduce the data, we fix $\Lambda_Q=1$ GeV for every heavy-quark flavors as shown in the figure and mentioned in Sect. 2. From those $\Lambda$ parameters, we have $( f_D, f_B, f_{D_s}, f_{B_s})=(207.54,208.13,262.56,262.39)$ MeV.  Hence, we have the following tendency from the present model:
\begin{equation}
\label{eq:TEND}
( f_D\approx  f_B\approx210\,\mathrm{MeV})\,\,\,\,<
\,\,\,\,( f_{D_s}\approx  f_{B_s}\approx260\,\mathrm{MeV}).
\end{equation}
These computed values are well compatible with other theoretical estimations including lattice QCD simulations. Detailed comparisons are given in Fig.~\ref{FIG2}.
\section{Summary and conclusion}
In the present work, we have investigated the weak-decay constants for the heavy PS-mesons consisting of the $(u,d,s,c,b)$ flavors, i.e. $ f_{D,B,D_s,B_s}$, using ExNLChQM. We list two important observations in the present work:
\begin{itemize}
\item We employ a phenomenological correction factor for the inclusion of the strange quark, compensating the $1/N_c$ corrections. By doing that, we obtain $ f_{D,D_s,B,B_s}=(207.53,\,262.56,\,208.13,\,262.39)$ MeV, which are qualitatively compatible with available experimental and theoretical values.
\item It is justified that the vacuum structure due to the heavy quarks is modified in comparison to that for the light quarks by seeing that $\Lambda_q\ne\Lambda_Q$. This tendency also signals the breakdown of the SU(5) flavor symmetry at the quark level, in addition to $m_q\ne m_Q$.
\end{itemize}
\section*{Acknowledgments}
The present report is prepared for the proceedings of the 20th International IUPAP Conference on Few-Body Problems in Physics (FB20), $20\sim25$ August 2012, Fukuoka, Japan. The author would like to thank A.~Hosaka for fruitful discussions on this subject. 

\end{document}